\documentclass[%
 aip,
 amsmath,amssymb,
 reprint,%
]{revtex4-1}

\usepackage{graphicx}% Include figure files
\usepackage{dcolumn}% Align table columns on decimal point
\usepackage{bm}% bold math
\usepackage[utf8]{inputenc}
\usepackage[T1]{fontenc}
\usepackage{mathptmx}
\usepackage{amsmath}
\usepackage{setspace}
\usepackage{amssymb}
\usepackage{bm}
\usepackage{epstopdf}
\setcitestyle{numbers,square}
\usepackage[usenames]{color}

\begin{document}

\preprint{AIP/123-QED}
\setcitestyle{square}

\title{{The zero-shear-rate limiting} rheological behaviors of ideally conductive particles suspended in concentrated dispersions under an electric field}
% Force line breaks with \\

\affiliation{Department of Mechanical and Materials Engineering, University of Nebraska-Lincoln, Lincoln, NE 68588-0526, USA}
\author{Siamak Mirfendereski}
\author{Jae Sung Park}
\email[Author to whom correspondence should be addressed; electronic mail:\\ ]{jaesung.park@unl.edu}

\date{\today}% It is always \today, today,

\begin{abstract}
The rheological behaviors of suspension of ideally conductive particles in an electric field are studied using large-scale numerical simulations {in the limit of zero-shear-rate flow}. {Under the action of an electric field}, the particles undergo the nonlinear electrokinetic phenomenon termed as dipolophoresis, which is the combination of dielectrophoresis and induced-charge electrophoresis. {For ideally conductive particles}, the dynamics of the suspension are {primarily} controlled by induced-charge electrophoresis. To characterize the rheological properties of the suspension, the particle stress tensor and particle pressure are calculated in a range of volume fraction up to almost random {close} packing. The particle normal stress and particle pressure are shown to behave non-monotonically with volume fraction, especially in concentrated regimes. In particular, the particle pressure is positive for volume fraction up to 30\%, after which it becomes negative, indicating a change in the nature of the particle pressure. The microstructure expressed by pair distribution function and suspension entropy is also evaluated. Visible variations in the local microstructure seem to correlate with the non-monotonic variation in the particle normal stresses and particle pressure. These non-monotonic behaviors are also correlated with the change in the dominant mechanism of particle pairing dynamics observed in our recent study [Mirfendereski \& Park, J. Fluid Mech. \textbf{875}, R3 (2019)]. Lastly, the effects of confinement on the particle stress and particle pressure are investigated. It is found that the particle pressure changes its nature very quickly at high volume fractions as the level of confinement increases. This study should motivate control strategies to fully exploit the distinct changing nature of the pressure for rheological manipulation of such suspension system. 
\end{abstract}

\maketitle

\begin{quotation}
%The ``lead paragraph'' is encapsulated with the \LaTeX\ 
%\verb+quotation+ environment and is formatted as a single paragraph before the first section heading. 
%(The \verb+quotation+ environment reverts to its usual meaning after the first %sectioning command.) 
%Note that numbered references are allowed in the lead paragraph.
%
%The lead paragraph will only be found in an article being prepared for the journal \textit{Chaos}.
\end{quotation}

\section{\label{sec:level1}Introduction}

An electric-field-driven suspension of particles in a viscous fluid has been widely studied in a number of fields including material science, micro- and nano-fluidics, and bioengineering \cite{bazant2010, sheng2012, nikonenko2014, li2018}. In general, there are two classes in such system. The most notable class is the so-called electrorheological (ER) suspension in which non-conductive but electrically active particles are suspended in an electrically insulating fluid. A wide spectrum of industrial applications for ER suspensions has been enabled by a rich rheological property \cite{wang2000}. The other class is comparatively new, where conductive or polarizable particles are suspended in an electrically conductive fluid or  {electrolyte}. Recently, such suspension has gained a growing interest in additive manufacturing or 3D printing technology and electrochemical energy storage technology \cite{presser2012, soloveichik2015, tan2019}. As highly concentrated suspensions are commonly used for these technologies, understanding the rheological behaviors of such suspensions is of practical interest. However, {the studies on the rheology} of such suspensions remain limited \cite{bauer2014, akuzum2018} to which this work is intended. The dynamics of such suspensions is  governed by the so-called induced-charge electrophoresis (ICEP) for ideally polarizable particles \cite{mirpark2019,park2010,saintillan2008}. Recently, such suspension displays non-trivial behaviors in concentrated regimes \cite{mirpark2019}.

{In the class of ER suspensions, electronic control of stress transfer tends to establish the concept of distinct fluid types. These suspensions are sometimes denoted as smart fluids, leading to various applications, including active shock absorbers, clutches, brakes, dampers, actuators, and artificial joints} \cite{sheng2012, hao2001}. The ER fluids under the action of an electric field are known to undergo dielectrophoresis (DEP) and exhibit a dramatic viscosity enhancement, which is reversible and can be controlled by the electric field \cite{pohlDEP, von2002, sheng2012}. This increase in viscosity, potentially leading to a transition to solid-state, is known as a consequence of a rapid formation of particle chains and columns along the applied field direction due to dipolar interactions between particles \cite{halsey1992,halsey1990,fossum2006,park2011}. These fibrillated chain structures are regarded as a typical feature of ER fluids and were first found by Winslow \cite{winslow1947}. Above a limiting volume fraction, these fibrillated structures, which span the electrode gap in quiescent suspensions, tend to become larger {with particle concentration} and eventually degrade with increasing shear rate in a shear flow \cite{parthasarathy1996}. It has also been well-known that the unique rheological properties of the ER fluids are attributed to the resistance to deformation of the field-induced structures \cite{bonnecaze1992dynamic,parthasarathy1996}. 

In a similar fashion, the presence of external magnetic fields in a suspension of highly magnetizable particles can also lead to a significant change in apparent viscosity and a reversible transition from liquid to nearly solid-state \cite{vicente2011, bossis1997}. Such suspension is also known as termed as a magnetorheological (MR) fluid. The viscosity of MR fluids, which contain the particles with a typical diameter of 1-10 $\mu m$, can also be tuned with field strength. As opposed to MR fluids, ferrofluids are composed of much smaller magnetic particles and exhibit a very small change in viscosity in the presence of an external field due to the considerable Brownian effects, disrupting the formation of the structures \cite{klingenberg2001}.

Rheological properties due to dipolar interactions have received much attention in both fields of electrorheology and magnetorheology \cite{martin1998, adriani1988}. Sim \textit{et al.} studied the large amplitude oscillatory shear (LAOS) behavior of ER fluids, relating the nonlinear LAOS behavior to the microstructural change of the suspension \cite{sim2003}. They also found that the cluster formation along with a slight rearrangement within a cluster results in the strain overshoot phenomenon, often seen in the system of complex fluids. Bonnecaze and Brady observed different types of rearrangements at small Mason numbers (Mn), the ratio of viscous to electrostatic forces \cite{bonnecaze1992dynamic}. They correlated the macroscopic rheology to the dynamics of the suspension for a range of Mn.
They also showed that the decrease in Mn, which is associated with an increase in electrostatic force relative to the viscous force, leads to an increase in effective viscosity \cite{bonnecaze1992yield, bonnecaze1992dynamic}.
% (I DON'T UNDERSTAND).
This Mason number is originally used for MR fluids, where it represents the ratio between the viscous and magnetic forces. Similar to ER fluids, the apparent shear viscosity of an MR suspension was seen to collapse onto a single function of Mn for a range of conditions \cite{dominguez2005,klingenberg2007}. %where the increase of Mn number, which is associated with a decrease in magnetic force relative to the viscous force, was also reported to reduce the apparent viscosity \cite{dominguez2005,klingenberg2007}.
 It is worth noting that Anderson claimed that the polarizability of particles could significantly affect the ER behavior of the suspension as conductive particles can acquire additional charges and in turn fail to participate in the formation of the structure as opposed to dielectric particles \cite{anderson1991}. 

As for the other class, which is the case of interest for this study, conductive particles are suspended in a conducting fluid. The dynamics of the system becomes more complex and far apart from that of the conventional ER suspensions governed by DEP interactions. In this system, the induced-charged electrophoresis (ICEP) \cite{squires2004, squires2006} arises as the particles acquire additional surface charges resulting in a nonlinear fluid flow around particle surfaces. {This phenomenon was first analyzed in the Russian colloidal literature \cite{gamayunov1986, dukhin1986}, which was reviewed by Murtsovkin \cite{murtsovkin1996}, and later rediscovered and coined by Squires and Bazant \cite{squires2004}.}
 It has been observed that ICEP is predominant over DEP {for a suspension of ideally conductive particles} \cite{park2010}, but this predominance can be modulated by surface contamination \cite{park2011d}. Recent numerical simulations were performed to investigate the effects of ICEP on the dynamics of such suspensions in a range of volume fractions up to a maximum of about 60\%, where non-trivial behaviors in large-scale dynamics were discovered in concentrated regimes (volume fraction of 45-50\%) and explained by the nature of contact mechanisms \cite{mirpark2019}. Yet, no rheological studies of such system have been done in the literature, which motivates the current study. 
 
It should be noted that owing to qualitatively similar far-field fluid disturbances governed by a Stokes dipole, the system studied shares similarities with active suspensions such as a suspension of spherical squirmers driven with a prescribed axisymmetric tangential velocity \cite{lauga2009, saintillan2018}. Hydrodynamic interactions in active suspensions induced by permanent swimming dipoles result in complex dynamics and mixing and diffusive behaviors \cite{saintillan2010}, which are qualitatively similar to ones of the current system governed by ICEP. However, the expected differences are the magnitude and orientation of the surface (or slip) velocity, which may modify the relative importance of attractive and repulsive interactions between particles \cite{ishikawa2006, evans2011}. In the context of the rheology of active suspensions, theoretical, numerical, and experimental studies have shown that particle activity causes an increase in intrinsic viscosity of the suspension of pullers, while reducing the intrinsic viscosity of the suspension of pushers when the particles are strongly active relative to the weak external flow (low P\'eclet numbers) \cite{saintillan2010,hatwalne2004, haines2008,haines2009}.
%, while it leads to decrease in that of pushers in weak external flows and strongly active suspensions relative to the flow (low P\'eclet numbers)(WHAT DO YOU MEAN BY WEAK FLOWS AND STRONGLY ACTIVE SUSPENSIONS?).the active contribution of pushers has been reported to 
% It is also consistent with the observations of the suspension of living algae (puller), where it becomes more viscous than the suspension of dead ones and the suspensions of both E. coli and sperms (pusher) are less viscous when alive than dead \cite{mcdonnell2015}.
It is also consistent with the observations that the suspension of living algae (pullers) is more viscous than the suspension of dead ones, while the suspension of both E. coli bacteria and sperms (pushers) is less viscous when alive than dead \cite{mcdonnell2015}.

In this paper, large-scale numerical simulations are used to investigate the rheology of the suspension of non-colloidal, ideally conductive spheres in a uniform electric field for volume fractions up to random {close} packing {in the limit of zero-shear-rate flow}. The suspensions under these conditions are known to undergo both induced-charged electrophoresis and dielectrophoresis. A combination of these two phenomena is also known as dipolophoresis (DIP) \cite{shilov1981}. The particle-particle interactions arising from DIP are generally governed by ICEP in a suspension of ideally conductive particles since the leading-order contributions to ICEP and DEP interactions are a Stokes dipole and a potential quadrupole, respectively, where the former has a slower decay with separation distance $R$ as $O(R^{-2})$ than the later as $O(R^{-4})$ \cite{saintillan2008, park2010, mirpark2019}. Under these conditions, the particles undergo random chaotic motions, which result in the constant rearrangement of the particle configurations but do not lead to the formation of chains as in the case of DEP only. The details of the simulation method adopted to facilitate the study of highly concentrated suspensions are reported in Sec.~\ref{sec:Simulation_method}. In Sec.~\ref{sec:result}, the simulation results for rheology and microstructure in the suspension as functions of volume fraction along with the system relaxation are presented and discussed. The effect of wall-confinement on the rheology is also evaluated in Sec.~\ref{sec:confined}.

\section{Governing equation and simulation method}\label{sec:Simulation_method}
We consider a suspension of $N$ identical neutrally buoyant ideally polarizable spheres of radius $a$ in a viscous electrolyte with the permittivity $\varepsilon$ and viscosity $\eta$. A cubic periodic domain is used to simulate an unbounded, infinite suspension. An external uniform electric field $\textbf{\textit{E}}_0 = E_0\hat{\mathbf{z}}$ is applied in the $z$ direction. %, particle-particle interactions arise as a result of DEP and ICEP, leading to relative motions.
The particles are assumed to carry no net charge, so the linear electrophoresis is not expected to occur. We also assume weak electric fields, thin Debye layers, and zero Dukhin number for no surface conduction \cite{squires2004}. It is also assumed that the particle size is large enough so that Brownian motion is negligible. Under these conditions and assumptions, the suspension dynamics results entirely from the effect of dipolophoresis.

{As a uniform electric field is applied, electric and hydrodynamic interactions between particles arise as a result of dielectrophoresis (DEP) and induced-charged electrophoresis (ICEP), which may lead to relative motions of particles. For ICEP, each spherical particle polarizes and forms a non-uniform surface charge distribution, which then attracts counter-ions in an electrolyte. The migration and accumulation of these counter-ions near the polarized surface result in the formation of a non-uniform Debye layer. The charging time of this non-uniform Debye layer is very small on the order of $\tau_c=\lambda_Da/D$, where $\lambda_D$ is the Debye layer screening length, $a$ is the particle radius, and $D$ is the characteristic diffusivity of ions in solution. % is very small.
 In a typical experiment ($a \sim 10$ $\mu$m, $\lambda_D \sim$ 10 nm, $D \sim 10^{-5}$ $\textrm{cm}^2$ s$^{-1}$), the charging time $\tau_c \sim 10^{-4}$ s, which is much smaller than the diffusion time across the particle $\tau_a = a^2/D \sim 10^{-1}$ s. Therefore, the non-uniform Debye layer can be assumed to remain at equilibrium. The effect of the electric field on the non-uniform Debye layer drives disturbance flows around the particle surface, which may lead to relative motions due to ICEP.}   
{The current calculation for ICEP interactions is based on the standard model of induced-charged electroosmosis (ICEO) around an ideally conductive sphere \cite{squires2004}, which has limitations compared to experimental observations, such as electrostatic correlations, electroviscous effects, dielectric decrement, among others \cite{bazant2009}. Thus, it should be noted that in general, the various neglected conditions are likely to modify the magnitude of ICEP interactions and shift their frequency response compared to the standard model used in the current study.} 
%\SM{In the current calculations, we use the standard model of induced-charged electroosmosis (ICEO) \cite{squires2004}, which neglects multiple phenomena such as electroviscous effect, the crowding of counter-ions, and electrostatic correlations, and so on \cite{bazant2009}. Thus, the model is based on the assumption that the bulk conductivity and diffuse-layer capacitance are constant, the applied voltage is small enough not to perturb the bulk salt concentration, and the local viscosity remains constant throughout the suspending fluid. Note that the current model has many limitations when compared with experimental results and may overestimate the magnitude of ICEO flow and fail to predict the actual frequency of response at certain conditions such as high concentration of ions and high applied voltage \cite{bazant2009}.}
{In addition, given that in the current model, Debye layers are sufficiently thin ($\lambda_D / a \approx 10^{-3}$) and the minimum distance between particles in a suspension is larger than $2\lambda_D$ even at high volume fractions, it can also be assumed that the Debye layer overlapping is negligible. Hence, the capacitance or polarizability of particles remains at equilibrium even at high volume fractions. However, in the case of highly concentrated suspensions, the Debye layer overlapping might happen due to strong excluded volume interactions, leading to the modification of the Debye layer capacity \cite{jonesem}, which is not considered in the current study. For DEP, the presence of other particles like a suspension causes disturbances to the local electric field around particles, resulting in a non-uniform Maxwell stress tensor $O(E^2)$ in the fluid. This tensor can yield a non-zero DEP force on surrounding particles, leading to relative motions in particle suspensions due to DEP \cite{park2010}.}

{In a uniform electric field, both ICEP and DEP do not lead to the motion of a single sphere owning to fore-aft symmetry. Indeed, the presence of other particles in a suspension leads to the symmetry breaking, resulting in the relative motion between the particles due to ICEP and DEP. The detailed description of paring dynamics associated with DIP for ideally conductive spheres in a uniform electric field was presented in the previous studies \cite{saintillan2008, park2010}.}

For electric and hydrodynamic interactions in a suspension, the method of simulation is based on the previous work of Park and Saintillan \cite{park2010}, for which a new approach was introduced to efficiently prevent particle overlaps in our recent work \cite{mirpark2019}. The outline of the method is as follows. Based on pair interactions due to DIP \cite{saintillan2008}, the translational velocity $\dot{\textbf{\textit{x}}}_\alpha$ of a given particle $\alpha$ in a suspension can be expressed by:
\begin{eqnarray}\label{eq:sums}
{\dot{\textbf{\textit{x}}}}_{\alpha} = \frac{\varepsilon a E_{0}^{2}}{\eta} \sum_{\beta=1}^{N} \left[\mathbf{M}^{DEP}(\textbf{\textit{R}}_{\alpha \beta}/a)+\mathbf{M}^{ICEP}(\textbf{\textit{R}}_{\alpha \beta}/a)\right] \boldsymbol{:}\hat{\textbf{\textit{z}}}\hat{\textbf{\textit{z}}}  \nonumber\\
  + M_0\mathbf{I}\boldsymbol{\cdot}\boldsymbol{F}^p   \quad\quad \alpha=1,...,N \ \ \ \ \ \ \
\end{eqnarray}
where $\textbf{\textit{R}}_{\alpha \beta} = \textbf{\textit{x}}_{\beta} - \textbf{\textit{x}}_{\alpha}$ is the separation vector between the particle $\alpha$ and particle $\beta$, and $\mathbf{M}^{ICEP}$ and $\mathbf{M}^{DEP}$ are third-order dimensionless tensors accounting for the ICEP and DEP interactions, respectively. {And $M_0$ is a mobility constant of a single sphere, $\mathbf{I}$ is an identity tensor, and $\boldsymbol{F}^p$ is the interparticle force for excluded volume interactions.} It is shown that these two tensors $\mathbf{M}^{ICEP}$ and $\mathbf{M}^{DEP}$ can be  entirely determined by the scalar functions of the dimensionless inverse separation distance $\lambda= 2a/\vert \textbf{\textit{R}} \vert$.  Specifically, for both DEP and ICEP, $\mathbf{M}$ is calculated as follows:
 \begin{equation}
   {\mathbf{M}}= 
   \left\{  
    \begin{array}{cc}
     {\mathbf{M}_p(\textbf{\textit{R}}/a)\ \ \ \ \ \ \ \ \ \ \ \ \ \ \ \ \ \ \ \ \ \ \ \ \ \ \ \ \ \ \ \ \ \ \ \ \ \ \ \ \ \ \ \ \ }  &  {{\textbf{\textit{R}}/a}\geqslant 4} \\
     {\mathbf{M}_p(\textbf{\textit{R}}/a)-\mathbf{M}_{FF}(\textbf{\textit{R}}/a)+\mathbf{M}_{TM}(\textbf{\textit{R}}/a)}  &  {\textbf{\textit{R}}/a< 4}   \\
    \end{array}
     \right . \label{eq:Mtotal}
 \end{equation}
where $\mathbf{M}_p$ denotes the periodic version of the far-field tensors $\mathbf{M}_{FF}$, which are given by:
\begin{align}
&\mathbf{M}_{FF}^{DEP}(\textbf{\textit{R}}/a)=\frac{1}{12}\mathbf{T}(\textbf{\textit{R}}/a)+O(\lambda^{5}), \label{eq:Mdepff}\\ 
&\mathbf{M}_{FF}^{ICEP}(\textbf{\textit{R}}/a)=-\frac{9}{8}\mathbf{S}(\textbf{\textit{R}}/a)-\frac{11}{24}\mathbf{T}(\textbf{\textit{R}}/a)+O(\lambda^{5}), \label{eq:Micepff}
\end{align}
where the two tensors
$\mathbf{S}$ and $\mathbf{T}=\nabla^{2}\mathbf{S}$ are the Green's functions for a Stokes dipole and for a potential quadrupole, respectively \citep{kim2013}. These two third-order tensors are also given in index notations as follows:
\begin{align}
&\displaystyle S_{ijk}(\textbf{\textit{R}})=-\frac{1}{R^{3}}(\delta_{ij}R_{k}-\delta_{ik}R_{j}-\delta_{jk}R_{i})-3\frac{R_{i}R_{j}R_{k}}{R^{5}}, \\
&\displaystyle T_{ijk}(\textbf{\textit{R}})=-\frac{6}{R^{5}}(\delta_{ij}R_{k}+\delta_{ik}R_{j}+\delta_{jk}R_{i})+30\frac{R_{i}R_{j}R_{k}}{R^{7}}.
\end{align}
These far-field tensors are asymptotically valid to order $ O(\lambda^4)$ for any pair of particles, and their use is consequently justified when they are sufficiently far apart (i.e., their separation distance is greater than $4a$). The far-field interactions ($\lambda \ll 1$) can be readily computed using the method of reflections. {These tensors reconfirm the dominance of hydrodynamic interactions due to ICEP over ones due to DEP in a suspension of ideally conductive particles}. However, the near-field corrections are necessary as the method of reflections becomes inaccurate when particles are close to each other (typically $|\textbf{\textit{R}}_{\alpha\beta}| < 4a$), for instance, during pairing events. This is achieved by correcting the far-field tensor $\mathbf{M}_{FF}$ with a more accurate version $\mathbf{M}_{TM}$ calculated using the method of twin multiple expansions \citep{saintillan2008}. This method is very accurate down to separation distances on the order of $|\textbf{\textit{R}}_{\alpha\beta}|\approx 2.005 a$ \citep{park2010}. {In addition, this method captures the strong modification of the local electric and hydrodynamic interactions between the two particles very accurately even at high volume fractions.}

To account for hydrodynamic interactions between particles $\alpha$ and $\beta$ and all its periodic images, the tensors of Eqs.~(\ref{eq:Mdepff}) and (\ref{eq:Micepff}) can be expressed as the periodic version of the far-field tensors, which are valid to order $O(R_{\alpha\beta}^{-4})$. Since a high-order computation $O(N^2)$ is required to direct calculation of the sums in Eq.~(\ref{eq:sums}), the smooth particle mesh Ewald (SPME) algorithm based on the Ewald summation formula of Hasimoto \cite{hasimoto1959} and on fast Fourier transforms is used to accelerate the calculation of the sums to $O(N \log N)$ operations \cite{park2010}. The details of the SPME algorithm on calculations of particle velocities can be found in Appendix A.

Once all the particle velocities $\dot{\textbf{\textit{x}}}_\alpha$ ($\alpha = 1, ..., N$) are calculated, particle positions are advanced in time using a second-order Adams-Bashforth time-marching scheme, with an explicit Euler scheme for the first time step. A fixed time step $\Delta t$ is used and is chosen so as to ensure that particles only travel a fraction of the mean inter-particle distance during one integration step. In order to prevent particle overlaps that occur due to the use of finite time steps in simulations, the application of a repulsive interparticle force is necessary. For this purpose, an effective algorithm functionally identical to the potential-free algorithm \cite{melrose1993} is implemented to prevent excessive particle overlaps as a result of DIP interactions, where particles are moved almost exactly, within roundoff errors $(\sim 2.005a)$, to contact. The form of the repulsive potential for excluded volume ($EV$) interactions is
\begin{equation}\label{eq:pfa}
   U^{EV}=\frac{1}{2}k\left ( 2a-R \right )^2,
\end{equation}
where $k$ is the time-step-dependent prefactor, which can be expressed as $k = 3\pi \eta a /\Delta t$ \citep{sherman2018}. {A particle velocity driven by a short-range repulsive force, which is the negative gradient of the potential with respect to the coordinates of the particle, can be obtained by the Stokes drag law.} {This repulsive force corresponds to the interparticle force in Eq. (\ref{eq:sums}).} The resulting {velocity} contributes to the displacement along the direction connecting the center of the two spheres at the points of closest approach. We tested the robustness with respect to the excluded volume interactions by using different values of the prefactor in Eq.~(\ref{eq:pfa}) and no excluded volume interactions, where almost identical results were produced. %It is found to be very successful in preventing the formation of infinite particle clusters even at a concentrated regime.
Another troublesome {factor} for suspension simulations in concentrated regimes is to provide the initial configuration of random particle distributions. To this end, the initial random configurations were generated using a similar procedure to ones suggested for dense hard-sphere systems \cite{rintoul1996computer,clarke1987numerical,stillinger1964systematic,jodrey1985computer}. In the present simulations, {all} runs were started with hard-sphere equilibrium configurations, but the first {100 time-steps} were discarded for better steady-state configurations beginning to compute averages.

In the remainder of the paper, all variables are made dimensionless using the following characteristic length and velocity scales: $l_{c}=a$ and $u_{c}=\varepsilon a E_{0}^{2} / \eta$.

\section{\label{sec:result}Results and discussion}
We performed large-scale simulations in a cubic periodic domain ($L_x \times L_y \times L_z = 20^3$) for a range of volume fraction $\phi$ up to almost random {close} packing ($\phi \approx 64\%$). In our previous study, the non-trivial suspension dynamics was observed in concentrated regimes, where the velocity fluctuation, hydrodynamic diffusivity, and number density fluctuation tend to increase with volume fraction at $\phi = 35$ before reaching a local maximum at $\phi\approx45\%$ and then drop as approaching to the random {close} packing  \cite{mirpark2019}. We attributed this non-monotonic behavior to the change in the dominant direction of particle-particle contacts from the field direction to the transverse direction. In this study, the local microstructure, which is correlated with the suspension dynamics, is evaluated by a pair distribution function. The rheology of the system is characterized by computing suspension bulk stress. Particle extra stress tensor $\Sigma_{p}$, which provides the essential explanation of the mean particle effect on the flow, is calculated based on the Batchelor's calculation of the average stress tensor in particle suspension \cite{batchelor1970}. The particle pressure as an isotropic part of the stress is evaluated for a range of volume fraction \cite{jeffrey1993}. 

\begin{figure}
\centerline{
\includegraphics[width=2.9in]{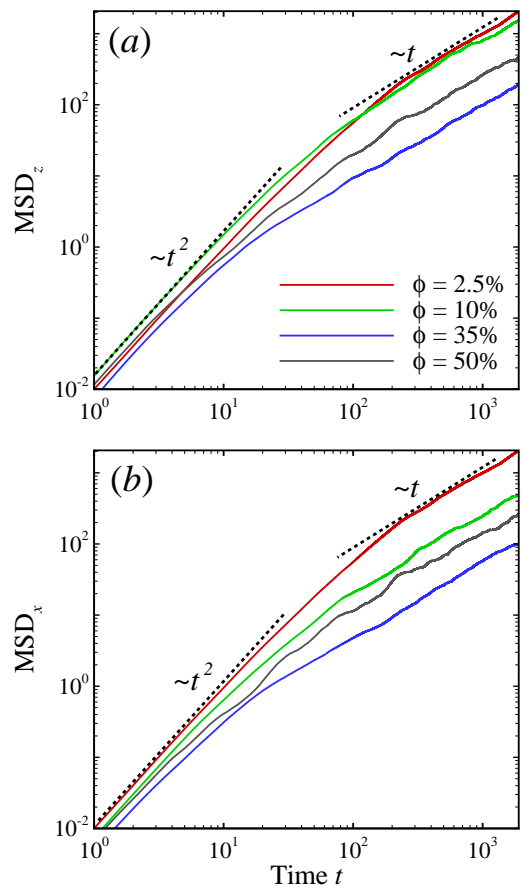}}% Here is how to import EPS art
\caption{\label{fig:MSD_AC1} The mean-square displacement (MSD) in ($a$) the $z$ direction (field direction) and ($b$) the $x$ direction (transverse direction) as a function of time on a log-log scale at four different volume fractions.}
\end{figure}
    
\begin{figure}
\centerline{
\includegraphics[width=2.9in]{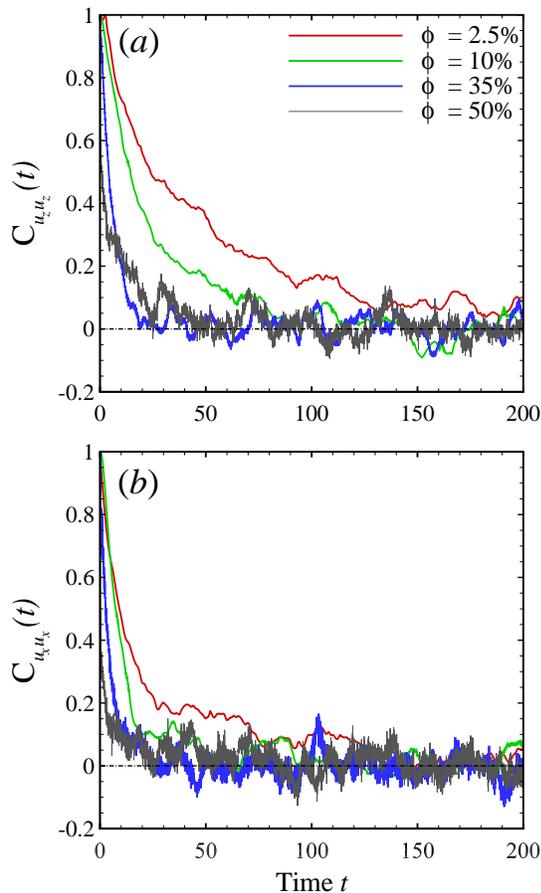}}% Here is how to import EPS art
\caption{\label{fig:MSD_AC2} The autocorrelation function of the particle velocity components in ($a$) the $z$ (field) direction and in ($b$) the $x$ (transverse) direction as a function of time for four different volume fractions.}
\end{figure}

\subsection{Suspension dynamics: relaxation time and interparticle distance}
The mean-square displacement (MSD) of particles versus time, which generally has been served as a start of the pathway to quantify the collective dynamics of suspended particles, is presented for four different volume fractions in Fig.~\ref{fig:MSD_AC1}($a$) and ($b$) for the $z$ direction (field direction) and the $x$ direction (transverse direction), respectively. After a few particle-particle interactions, the initial quadratic growth of the MSD curve is followed by a transition to the diffusive regime in which the MSD curve linearly grows with time in both $x$ and $z$ directions. To compute a time required to reach the diffusive regime due to loss of memory, denoted as a relaxation time (also known as crossover time), the autocorrelation functions $C_{uu}$ of particle velocities in the $x$ and $z$ directions are calculated, as seen in Fig.~\ref{fig:MSD_AC2}. The relaxation time can be approximated as the time when the function reaches its first global minimum. In both directions, the function decorrelates very quickly with increasing volume fraction.
%It is shown that the relaxation time seems to decrease until $\phi=35\%$ and then increase with volume fraction at $\phi=50\%$. 

\begin{figure}
\centerline{
\includegraphics[width=3.0in]{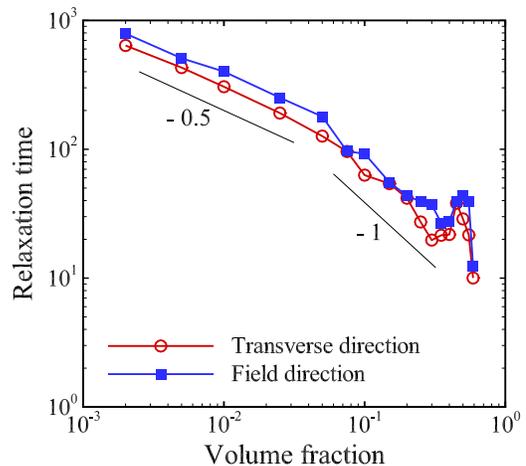}}% Here is how to import EPS art
\caption{\label{fig:RT} The relaxation time in the transverse ($x$) and field ($z$) directions as a function of volume fraction on a log-log scale. Different zones associated with the different slopes are seen on a log-log scale for $\phi < 35\%$, and a non-trivial variation in relaxation time is seen for $\phi > 35\%$.}
\end{figure}

Fig.~\ref{fig:RT} shows the relaxation time $\tau$ as a function of volume fraction $\phi$ on a log-log scale. In the dilute regime $(\phi < 5\%)$, the relaxation time tends to decrease with volume fraction and scale as $\tau\sim\phi^{-0.5}$. Subsequently, in the semi-dilute regime ($\phi=5\%-35\%$), the relaxation time decreases faster than for the dilute regime, approximately scaling as $\tau\sim\phi^{-1}$. The decreasing trend in the relaxation time with volume fraction is easily explained by an increase in the magnitude of particle-particle interactions with increasing volume fraction. %The faster reduction in the separation distance is likely to result in faster transition to a diffusive state. Interestingly, the separation distance decreases even faster at concentrated regimes ($\phi > 35\%$), but 
Interestingly, the relaxation time appears to increase at $\phi\approx 35\%$ up to $\phi \approx 47.5\%$ and then decrease again as approaching random {close} packing. {It is} this range of volume fraction that the non-trivial behavior of increasing hydrodynamic diffusivity and velocity fluctuation was observed in our previous study for the same suspension \cite{mirpark2019}. %Therefore, similar non-trivial behavior can also be observed in relaxation time, which seems to increase with volume fraction in concentrated regime, and then drops as approaching the random packing fraction. %It suggests that the change in hydrodynamics diffusivity in concentrated suspension is associated with change in relaxation time.     

\begin{figure}
\centerline{
\includegraphics[width=3.0in]{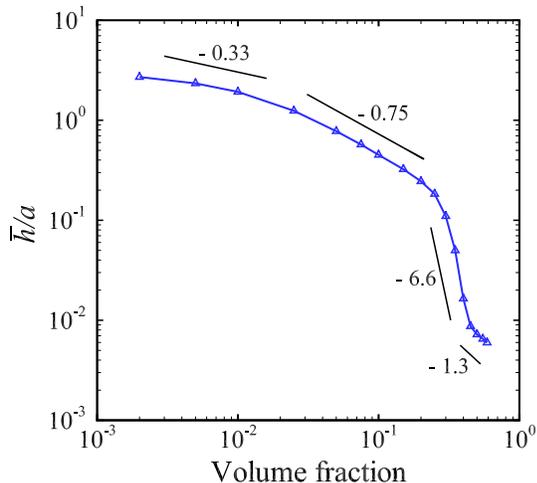}}% Here is how to import EPS art
\caption{\label{fig:dist} The average interparticle gap $\bar{h}/a = \bar{R}_{min}/a-2$ as a function of volume fraction on a log-log scale. Four different zones associated with different slopes in a power law of the volume fraction are identified.}
\end{figure}

To further distinguish different behaviors in different ranges of volume {fraction}, the mean interparticle gap $\bar{h} = \bar{R}_{min}-2a$ is calculated as a function of volume fraction, as shown in Fig.~\ref{fig:dist}. Four different zones can be readily distinguished by different slopes in the log-log plot, namely dilute, semi-dilute, concentrated, and very concentrated (close to random {close} packing) regimes. In dilute regime (first zone), the average interparticle gap is proportional to $\bar{h} \sim {\phi}^{-1/3}$, which has also been observed in many dilute suspensions \cite{hao2011, michot2006, jiang1998, borggreve1987}. The exponent then becomes larger at $\phi \approx 5\%$, providing the second slope of $-0.75$ (second zone). In concentrated regime (third zone), it decreases dramatically with {a} much higher exponent as a volume fraction increases. Finally, the slope in the very high concentrated regime (fourth zone) decreases slowly compared to the third zone because the average interparticle gap is approaching the value of designated prefactor corresponding to excluded volume interaction in Eq.~(\ref{eq:pfa}).
  
\subsection{\label{sec:level2}Suspension microstructure and {entropy}}
\begin{figure*}
\centerline{
\includegraphics[width=8.0in]{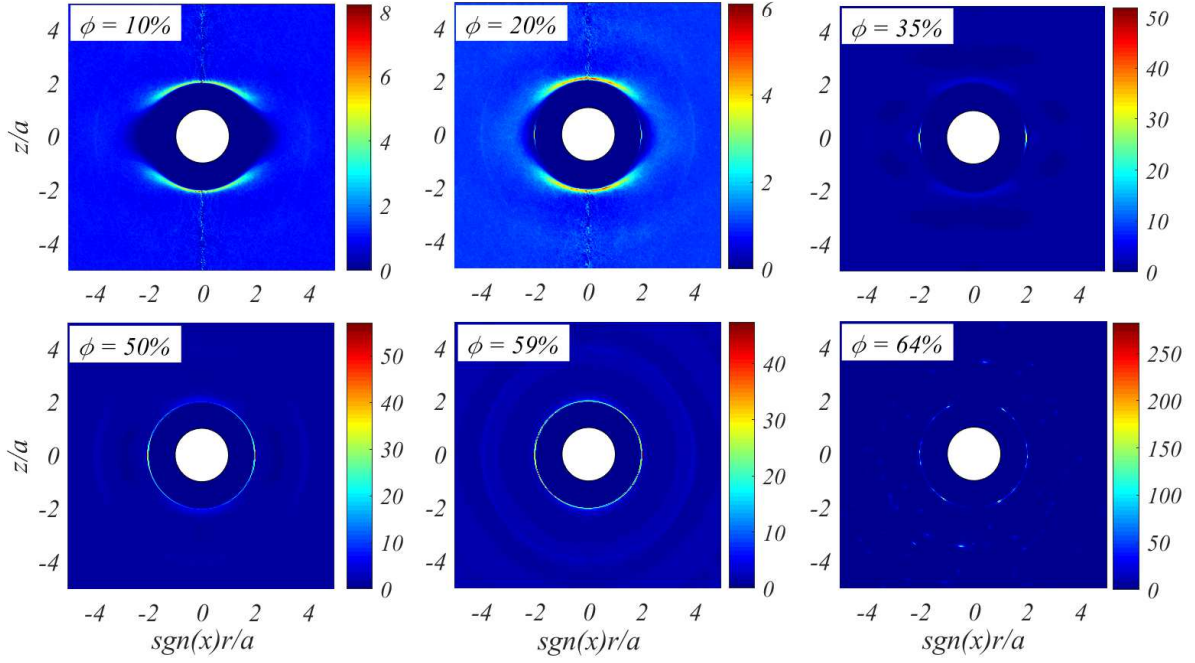}}% Here is how to import EPS art
\caption{\label{fig:pdf} Pair distribution functions in suspensions undergoing dipolophoresis at six different volume fractions in coordinates ($sgn(x)r,z$), where $r^2=x^2+y^2$. A probe particle is located at the origin (white areas are indeed an excluded volume).}
\end{figure*}

The suspension microstructure is known to provide important information in complex flows, especially its implications to rheology \cite{stickel2005,morris2009,butler2018}. An appropriate way to characterize the suspension microstructure, which is the spatial arrangement of particles, is to calculate the pair distribution function. This function provides information about the probability of finding a particle with respect to a probe one at the origin. Fig.~\ref{fig:pdf} shows the effect of volume fraction on this function. As seen in the figure, the maximum probability is first located near the particle poles at $\phi = 10\%$ and $\phi = 20\%$, as previously observed in both suspension of spherical \cite{park2010, mirpark2019} and rod-like \cite{saintillan2006} particles. In the concentrated regime ($\phi > 30\%$), the maximum shifts from the poles to equators due to the change of the dominant mechanism of pairing dynamics \cite{mirpark2019}. With a further increase in volume fraction up to $\phi = 59\%$, this shifted maximum probability region seems to emanate and propagate over the particle surface toward the poles  where the peak region finally turns to entirely cover the particle surface, forming seemingly microstructural isotropy. {Specifically, starting at $\phi \approx 20\%$, the dominant mechanism and direction of particle parings change from attractive in the field direction to repulsive in the transverse direction due to strong ICEO flows in the lateral direction. This change in the dominant mechanism leads to pushing the particles to contact in the transverse direction, resulting in the appearance of the second high probability region at the equator [11]. The further increase in the repulsive interactions with volume fraction eventually causes nearly 95\% of particle contacts to occur in the lateral directions, leading to a clear transition of the high probability region from the polar to the equator at $\phi \approx 35\%$. As a volume fraction is further increased, the increase in excluded volume interactions results in increasing particle contacts in all directions by which the high probability region starts to entirely cover particle surface for $45\%\leqslant \phi \leqslant 59\%$}. {Again, it should be noted that the local microstructure of the current system is primarily governed by ICEP as the pair distribution function of the suspensions undergoing only ICEP (not shown) is almost identical to that of Fig.~\ref{fig:pdf}.} Finally, the function seems to show a crystal structure at random {close} packing ($\phi = 64\%$), where the multiple high probability spots indicates the typical crystal structural information \cite{sierou2002,foss2000,gasser2010}. %The drastic change in the microstructure probably ties to the non-trivial dynamics, implies distinctive system rheology that we will address in Sec.~\ref{sec:level1}3}

\begin{figure}
\centerline{
\includegraphics[width=3.0in]{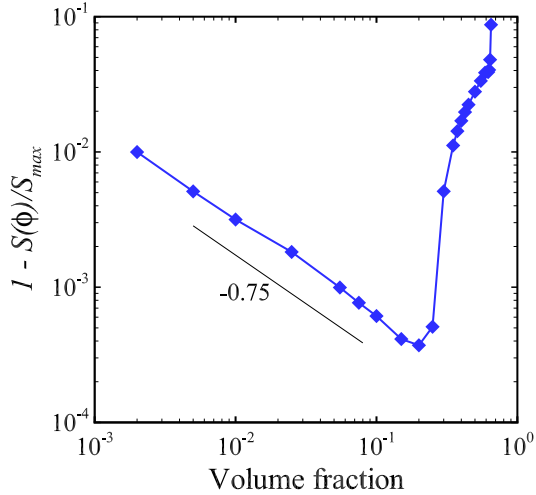}}% Here is how to import EPS art
\caption{\label{fig:ent} The normalized deviation of the suspension entropy from the maximum entropy ($1-S(\phi)/S_{max}$) as a function of volume fraction on a log-log scale, where $S(\phi)$ is the entropy of suspension at a certain volume fraction $\phi$ and $S_{max}$ is the maximum entropy resulted from the equiprobable pair distribution function. }
\end{figure}

Having determined the pair distribution functions, the suspension microstructure can be linked to the suspension entropy. Typically, the Shannon entropy $S$ is calculated \cite{shannon1948} and is given by
\begin{equation}
   S(\phi)=-\sum_{i=1}^{M}P(x_{i})\textrm{log}_2P(x_{i}), \label{eq:ent}
\end{equation}
where $P$ is the pair distribution function at a volume fraction $\phi$, $x_i$ is the $i$th location in $P$, and $M$ is the total number of the locations in $P$. This entropy is also known to quantify the order of a suspension. Fig.~\ref{fig:ent} shows the deviation of the suspension entropy normalized by the maximum entropy, $1-S/S_{max}$, as a function of volume fraction. The maximum entropy, $S_{max}$, results from an equiprobable pair distribution function, meaning that the probability of finding a particle is constant at all locations.
 It is found that the normalized suspension entropy {($1-S/S_{max}$)} decreases as $\sim \phi^{-0.75}$ from $\phi=0.25\%$ to $\phi=20\%$ {at which the Shannon entropy $S$ eventually reaches a maximum value, meaning} that the suspension becomes the least ordered. The normalized entropy starts to increase and reach the maximum value at the random {close} packing, where the crystal structure is likely to be formed. At the volume fraction of the minimum normalized entropy ($\phi = 20\%$), the secondary high probability region starts to appear at the equators in a pair distribution function (see Fig.~\ref{fig:pdf}). In addition, it is worth noting that at this volume fraction, the nature of pairing dynamics is changed from repulsive to attractive, where particle contacts along the transverse direction become predominant over the attractive contacts along the field direction \cite{mirpark2019}. 

\subsection{\label{sec:level3}Particle stress}
For defining the rheological properties of the suspension, the calculation of the bulk stress $\left \langle \boldsymbol{\Sigma} \right \rangle$ is needed, where the angle bracket denotes an ensemble average over the particles. Here, we use the Batchelor's calculation for the average stress tensor in a suspension of force-free particles \cite{batchelor1970}. The average of bulk stress is then determined by the average of the Cauchy stress tensor $\boldsymbol{\sigma}$ over a characteristic volume $V$: 
\begin{equation}
\left \langle \boldsymbol{\Sigma} \right \rangle =  \frac{1}{V} \int_V {\boldsymbol\sigma}(\boldsymbol{x})dV = -p_f \mathbf{I} + 2\eta \left \langle \mathbf{E}^\infty \right \rangle +  \boldsymbol{\Sigma}^p , \label{eq:bstress}
\end{equation}  
where $p_f = \langle p \rangle_f$ is the (constant) pressure in the fluid, $\mathbf{I}$ is the identity matrix, and $2\eta\langle\mathbf{E}^\infty\rangle$ is the deviatoric contribution from an incompressible fluid containing rigid particles. These first two terms are the contributions of the fluid to the stress, comprising a Newtonian behavior \cite{sierou2002}. The non-Newtonian behavior is captured by the particle contribution to the stress ${\boldsymbol\Sigma}^p$, which is given by (in index notation)
\begin{equation}\label{eq:pstress1}
\begin{aligned}
{\Sigma}^p_{ij}  &= \frac{1}{\mathit{V}}\sum_{\alpha=1}^N \int_{A_0}\left [\sigma_{ik}x_jn_k^{\alpha} - \eta (u_in_j^{\alpha} + u_jn_i^{\alpha})  \right ]d{\mathit{A}}\\
&= \frac{1}{\mathit{V}}\sum_{\alpha=1}^N S_{ij}^{\alpha} +  \frac{1}{\mathit{2V}}\sum_{\alpha=1}^N \epsilon_{ijk} L_{k}^{\alpha},
\end{aligned}
\end{equation}
where $\mathit{A}_0$ is the surface of particle $\alpha$, $u_i$ is the velocity component in an ambient fluid, and ${n^{\alpha}_i}$ is the component of the normal vector pointing outward from the particle surface to the fluid. $\boldsymbol{S}^{\alpha}$ and $\boldsymbol{L}^{\alpha}$is the stresslet and torque on particle $\alpha$, respectively. For torque-free rigid particles, in which case the present suspension is, the stresslet is the only contribution to the particle stress and is simply computed by 
\begin{equation}\label{eq:stresslet}
S_{ij}^{\alpha} = \int_{A_0}\left [\frac{1}{2}(\sigma_{ik}x_j + \sigma_{jk}x_i)n_k^{\alpha} \right ]d{\mathit{A}},
\end{equation}
where the position vector $\mathbf{x}$ can be taken from an arbitrary origin as long as the total hydrodynamic force on $N$ particles vanishes. The particle stress tensor is then readily obtained by
\begin{equation}\label{eq:pstress2}
{\Sigma}^p_{ij} = n\left \langle S_{ij} \right \rangle,
\end{equation}
where $n = N/V$ is the number density of the suspension. Specifically, the particle stress is the number density times the average symmetric force dipole per particle.

For the suspension considered, which includes non-Brownian particles in a quiescent flow, the stresslet can be further decomposed into two stress-generating mechanisms, namely ICEP effects and hydrodynamic interactions
\begin{equation}\label{eq:pstress3}
{\Sigma}^p_{ij}  = n\left \langle S_{ij}^{ ICEP} \right \rangle \ + n\left \langle S_{ij}^H \right \rangle.
\end{equation}
The first term is associated with an induced nonlinear slip velocity over particle surface due purely to electric interactions, while the second term corresponds to the effect of hydrodynamic interactions between particles. Note that that as mentioned above, the effect of electrostatic forces (dielectrophoresis effect) is negligible compared with induced-charged electrophoresis for the system studied; therefore, the contribution of DEP to the stress would not be included. %{Additionally, the contribution of a short-range repulsive force to the bulk stress is negligible because it decays very quickly to zero within a very short distance $h \ll a$ \cite{nazockdast2012}.}

\begin{figure}
\centerline{
\includegraphics[width=3.1in]{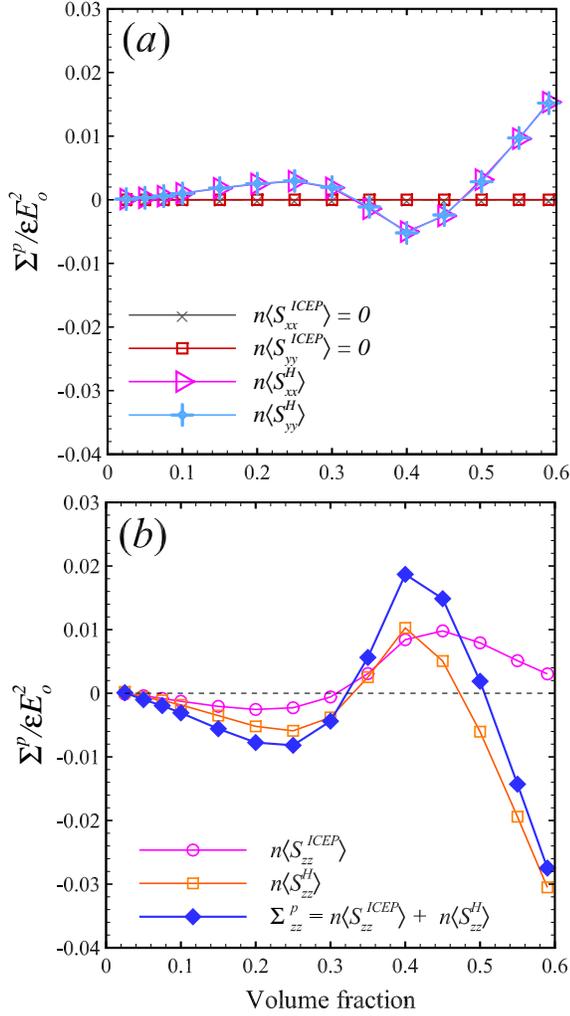}}% Here is how to import EPS art
\caption{\label{fig:nstress} Dependence of the particle normal stresses, namely ICEP ${\left \langle S^{ICEP}\right \rangle}$ and hydrodynamic $\left \langle S^{H}\right \rangle$ contributions, on volume fraction in ($a$) the transverse ($x, y$) directions (note that $\Sigma^p_{xx}=n{\left \langle S^{H}_{xx}\right \rangle}$ and $\Sigma^p_{yy}=n{\left \langle S^{H}_{yy}\right \rangle}$) and ($b$) the field ($z$) direction. ICEP effect only contributes to the particle normal stress in the field direction. Due to symmetry with respect to the field direction, $\Sigma^p_{xx}=\Sigma^p_{yy}$. }
\end{figure}

As the slip velocity on the particle surface drives flow in a suspending fluid, the velocity field around a given particle due to the slip velocity can be found by solving the Stokes equations \cite{squires2004, saintillan2008}. To leading order of $O(R^{-3})$, 
%it can be done as if the particle is isolated and 
the contribution of the slip velocity to the particle stress ${S^{ ICEP}_{ij}}$ is now given by
\begin{equation}\label{eq:s_ICEP}
S^{ICEP}_{ij} = \frac{9}{16} \epsilon E^2_0  \left( \frac{a}{R}\right )^{3} (\textbf{\textit{A}} \hat{\textbf{\textit{E}}}_0 + \hat{\textbf{\textit{E}}}_0\textbf{\textit{A}}) + O(R^{-4}),
\end{equation}
where $\textbf{\textit{A}} \equiv \left ( \textbf{\textit{I}} - 3 \hat{\textbf{\textit{R}}} \hat{\textbf{\textit{R}}} \right ) \cdot \hat{\textbf{\textit{E}}}_0 $, $\hat{\textbf{\textit{R}}} = {\textbf{\textit{R}}}/a $, and $\hat{\textbf{\textit{E}}}_0 = {\textbf{\textit{E}}}_0/E_0$. Subsequently, the velocity field generated by the slip velocity causes the motion of other particles, which can be obtained by Faxen's law \cite{kim2013}. This particle motion corresponds to the first effect of hydrodynamic interactions. The contribution of this hydrodynamic interaction to the particle stress ${S^H_{ij}}$ is then captured by the disturbance velocity field due to the particle motion when placed in the velocity field generated by the slip velocity. In other words, the contribution to ${S^H_{ij}}$ is related to the hydrodynamic interactions captured by the reflection of a velocity disturbance from one particle to another. To leading order of $O(R^{-3})$ in the first effect of hydrodynamic interactions, the contribution to ${S^H_{ij}}$ is given by
\begin{equation}\label{eq:sh_1}
\begin{aligned}
S^{H}_{ij} = \frac{45}{32}\epsilon E^2_0 \left( \frac{a}{R}\right)^{3} &[ \left ( \textbf{\textit{I}} - 3 \hat{\textbf{\textit{R}}}\hat{\textbf{\textit{R}}} \right ) - 3(\hat{\textbf{\textit{E}}}_0\cdot \hat{\textbf{\textit{R}}}) 
(\hat{\textbf{\textit{E}}}_0\hat{\textbf{\textit{R}}} + 
\hat{\textbf{\textit{R}}}\hat{\textbf{\textit{E}}}_0) \\
& - 3(\hat{\textbf{\textit{E}}}_0\cdot \hat{\textbf{\textit{R}}})^2 \left ( \textbf{\textit{I}} - 5 \hat{\textbf{\textit{R}}}\hat{\textbf{\textit{R}}} \right) ] + O(R^{-4}).
\end{aligned}
\end{equation}

Given these two contributions to the particle stress, we turn to the dependence of the particle stress tensor on a volume fraction. Fig.~\ref{fig:nstress} shows the diagonal entities (i.e., normal stresses) in the particle stress tensor. It was found that the diagonal entities are several orders higher in magnitude than non-diagonal ones (i.e., shear stresses), suggesting that the rheological behavior of the current system is essentially governed by the normal stresses.

Fig.~\ref{fig:nstress}($a$) shows the normal stresses in the transverse directions ($x$ and $y$ directions). To leading order, the ICEP effect appears to provide no contribution to the normal stresses in both $x$ and $y$ directions, essentially $S_{xx}^{ICEP} = S_{yy}^{ICEP} = 0$. However, the hydrodynamic interactions contribute to the particle normal stress in the transverse directions, which are almost the same in both $x$ and $y$ directions, that is $S_{xx}^{H} \approx S_{yy}^{H}$. %Thus, the stress system is symmetric with respect to the $x$ and $y$ directions as expected for a periodic domain with an electric field in the $z$-direction.
Further, non-trivial behavior is observed in terms of sign and trend. The transverse hydrodynamic normal stresses are positive in dilute regimes and increase with volume fraction before reaching a local maximum at $\phi = 25\%$. They start to decrease and become negative at $\phi \approx 32.5\%$, reaching a minimum at $\phi = 40\%$. They start to increase again and become positive at $\phi = 47.5\%$. Finally, they increase rapidly as approaching random {close} packing.

For the field direction, the particle normal stress is presented in Fig.~\ref{fig:nstress}($b$). As opposed to the transverse directions, the ICEP effect contributes to the particle normal stress in the field direction. Non-trivial behaviors are also observed in the field direction. However, in this case, both ICEP and hydrodynamic contributions are all negative in dilute regimes and become more negative with volume fraction up to $\phi = 25\%$. They then start to increase and become positive at $\phi \approx 32.5\%$, reaching a maximum at $\phi = 45\%$ and $\phi = 40\%$ for ICEP and hydrodynamic contributions, respectively. While the ICEP contribution stays positive for the rest of volume fraction with a slight decrease toward random {close} packing, the hydrodynamic contribution rapidly decreases and becomes negative again at $\phi = 47.5\%$, reaching a minimum at random {close} packing. {This} non-monotonic behavior of the particle normal stresses with volume fraction could imply distinctive rheological behaviors of the system studied.

Interestingly, for the hydrodynamic contributions, the normal stresses in the transverse directions, ${\left \langle S_{xx}^{H}\right \rangle}$ and $\left \langle S_{yy}^{H}\right \rangle$, are exactly opposite to that in the field direction, $\left \langle S_{zz}^{H}\right \rangle$.
 %This can be explained by the particle pairing mechanism originated due to DIP or ICEP, where particles are attracted along the field direction and separated each other along the transverse directions\cite{saintillan2008,park2010,mirpark2019}.
This can be explained by the stress-generating response of the particle pairing originated due to DIP or ICEP. Specifically, for each pair of particles undergoing ICEP, $S_{zz}^{H} = -(S_{xx}^{H}+S_{yy}^{H})$ always holds{. Considering} ${\left \langle S_{yy}^{H}\right \rangle} \approx {\left \langle S_{xx}^{H}\right \rangle}$ for the current system with an electric field applied in the $z$ direction{, then }${\left \langle S_{zz}^{H}\right \rangle}\approx-2{\left \langle S_{xx}^{H}\right \rangle}$.
Note that the particle pairing dynamics, where particles tend to be attracted along the field direction and repulsive along the transverse direction, essentially results in exactly opposite flow fields around particles as the particles move in the transverse and field directions, resulting in the opposite signs for stresslets \cite{saintillan2008,park2010,mirpark2019}.
  
The sign of all the contributions in the particle normal stress seems to be correlated with the local microstructure seen in Fig.~\ref{fig:pdf}. In dilute regimes, the high probability region in the pair distribution function is located near the particle poles, which corresponds to a negative sign in $\left \langle S_{zz}^{H}\right \rangle < 0$ and $\left \langle S_{zz}^{ICEP}\right \rangle < 0$, indicating that the particles are mostly paired up along the field direction (attractive pairings). Note that there is a positive sign in $\left \langle S_{xx}^{H}\right \rangle > 0$, but its effect is negligible on the microstructure. Their signs eventually change with the increase in volume fraction at $\phi \approx 32.5\%$, which corresponds to transition in local microstructure that the high probability regions now shift to the equators (repulsive pairings). With further increase in volume fraction, the second transition in particle normal stress associated with a change in the sign of hydrodynamic contributions is also confirmed by the local microstructure, where the high probability region in the equators propagates toward the poles. Therefore, it is suggested that the non-monotonic variation of the particle stress system ties into the variation of the microstructure due to a change in the dominant mechanism and direction of particle parings \cite{mirpark2019}.

{It is worth mentioning that the contribution of the short-range repulsive force to the bulk stress can be negligible because it decays very quickly to zero within a very short distance $h\ll a$ \cite{nazockdast2012}. To validate that the contribution of the short-range repulsive force to the bulk stress is negligible for the current system, we have calculated the hard-sphere interparticle stresslet $\langle S^P\rangle = -n\langle\boldsymbol{x}\boldsymbol{F}^P\rangle$ \cite{singh2000} in the range of volume fraction (not shown). It is seen that the maximum contribution of the short-range repulsive force to the particle normal stress is of $O(10^{-4})$, which is almost two orders of magnitude smaller than that of ICEP and hydrodynamic contributions shown in Fig.~\ref{fig:nstress}.}

\begin{figure}
\centerline{
\includegraphics[width=3.2in]{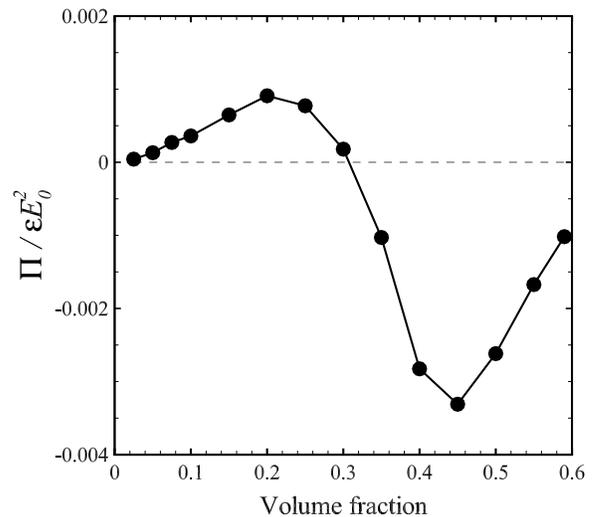}}% Here is how to import EPS art
\caption{\label{fig:ppressure} The particle pressure $\Pi$ as a function of volume fraction. The nature of particle pressure is changed at $\phi \approx 30\%$ as it turns into negative from positive.}
\end{figure}

\subsection{\label{sec:level4}Particle pressure}
The particle pressure represents the isotropic contribution of particles in bulk stress and is mechanically defined as the negative mean normal stress exerted by the particles
% through hydrodynamic interactions 
in a viscous fluid, i.e., $\Pi =  -1/3\Sigma^p_{ii}$ \cite{jeffrey1993}. The particle pressure is also referred to as the non-equilibrium continuation of osmotic pressure \cite{yurkovetsky2008, deboeuf2009}. This particle pressure has been employed in many studies to characterize the rheological properties of the suspension of hard particles under shear flow \cite{guazzelli2018,sierou2002,yurkovetsky2008}.
% (WHAT DOES IT MEAN? Although a suspension, the mixture of hard particles and a viscous fluid, is typically incompressible, the particle phase is not considering a macroscopic view. Therefore, particle pressure indeed plays a role in variation in particle phase volume.)
Although the whole suspension, which is the mixture of hard particles and a viscous fluid, is typically incompressible, the entire collection of the suspended particles is compressible from a macroscopic point of view \cite{guazzelli2018}. With that said, the particle pressure has been quantified based upon the tendency of particle phase to expand \cite{deboeuf2009}. As a way to evidencing the particle pressure, a U-shaped tube was sheared in a Couette device with a semi-permeable membrane placed in the middle to separate a pure liquid on one side and a suspension on another (i.e., the particles are restricted to pass the membrane) \cite{deboeuf2009}. The pure liquid has been observed to be sucked into the other side of a tube containing particles as the device is sheared. This observed suction indicates that the particle phase tends to expand, which means placing the suspending fluid in tension relatively, leading to positive particle pressure \cite{yurkovetsky2008}. In this regard, we investigate the sign of the particle pressure in the suspension studied. 

Fig.~\ref{fig:ppressure} shows the particle pressure as a function of volume fraction. To leading order of $O(R^{-3})$, it turns out that there is no hydrodynamic contribution to the particle pressure because $\left \langle S_{zz}^{H}\right \rangle = -(\left \langle S_{xx}^{H}\right \rangle + \left \langle S_{yy}^{H}\right \rangle)$. The only contribution to the particle pressure results from the ICEP effect. At $\phi \leqslant 30\%$, the particle pressure is positive ($\Pi > 0$) similar to that of the hard-sphere suspensions in a simple shear flow, which is always positive \cite{yurkovetsky2008,deboeuf2009}. Again, it implies that the particle pressure places the suspending fluid in tension relative to the pure fluid state as the particles exhibit volumetric expansion regarding $p_f=-\Pi$. Interestingly{,} the particle pressure becomes negative at $\phi > 30\%$, which indicates the change in its nature in a way that the suspending fluid is now relatively placed in compression as particles show volumetric contraction. Eventually, the particle pressure reaches its minimum and also maximum in magnitude at $\phi = 45\%$. It is at this volume fraction that a non-trivial local maximum was observed in suspension dynamics, for instance, hydrodynamic diffusivity and velocity fluctuation \cite{mirpark2019}. It suggests a clear relationship between the particle pressure and the suspension dynamics in concentrated regimes, %as the non-trivial behavior is also seen in the particle pressure (Fig.~\ref{fig:ppressure}).
where the negative particle pressure might be the manifestation of the change in the predominance of particle pairings from mild attractive contacts to massive, strong repulsive contacts \cite{mirpark2019}. %It is worth noting that $\boldsymbol{\nabla \cdot {\Sigma^p}} = 0$ for all volume fractions, indicating no particle migration on average in a suspension.

\begin{figure}
\centerline{
\includegraphics[width=3.2in]{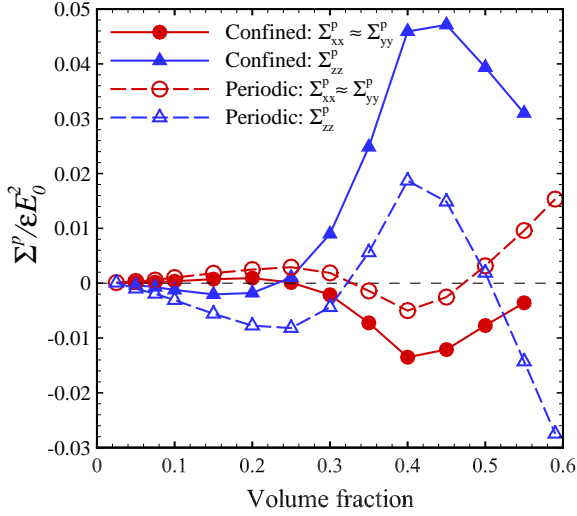}}% Here is how to import EPS art
\caption{\label{fig:psconfined} The particle normal stresses in the transverse ($x$) and field ($z$) directions for periodic and confined suspensions. The confinement is placed in the $z$ direction and the confinement level is $\chi = 0.1$, where $\chi = 2a/L_z$, $L_z = 20$, and $L_z$ is the electrode spacing. For both periodic and confined suspensions, $\Sigma^p_{xx} \approx \Sigma^p_{yy}$. }
\end{figure} 

\subsection{\label{sec:confined}Confinement effect}
It has been well-known that the hydrodynamics and rheology of suspensions are strongly affected by confinement \cite{swan2011,jeanneret2019,ramaswamy2017}. We investigate the effects of confinement on particle stress and pressure in the current system. To this end, we modify the boundary conditions in the $z$ direction from periodic to wall-confined conditions as followed by our previous study \cite{park2011}, where only short-range interactions with the boundaries were captured.
{Specifically, to introduce the rigid boundaries in a domain, a short-range repulsive force is imposed between the particles and the boundaries by implementing the algorithm similar to the one used to prevent excessive particle overlaps.} {However, it should be noted that long-range interactions with the boundaries are likely to have an effect on particle dynamics, particularly in the direct vicinity of the boundaries, which will be included in future work.} We introduce the confinement factor $\chi = 2a/L_z$ to quantify the level of confinement, where $L_z$ is the electrode spacing. The larger $\chi$, the stronger the level of confinement. Typically, $\chi < 0.05$ corresponds to a weakly confined regime \cite{graham2011}.

%To further characterize the effect of confinement on the particle stress and pressure,
\begin{figure}
\centerline{
\includegraphics[width=3.2in]{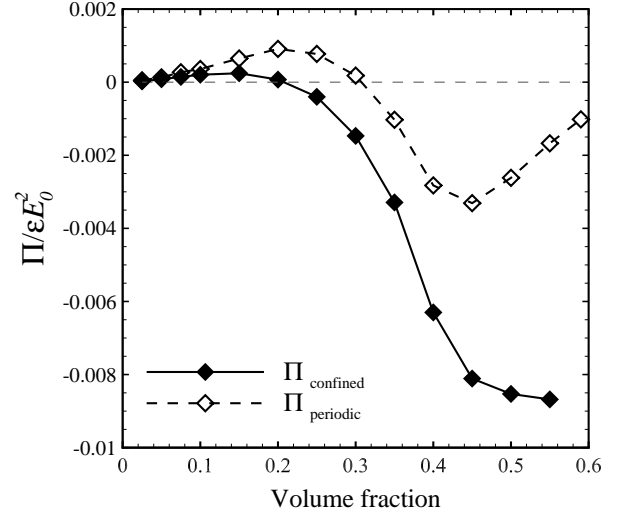}}% Here is how to import EPS art
\caption{\label{fig:ppconfined} The particle pressure $\Pi$ for periodic and confined suspensions. The confinement level is $\chi = 0.1$, where the confinement factor $\chi = 2a/L_z$, $L_z = 20$, and $L_z$ is the electrode spacing.}
\end{figure} 

Fig.~\ref{fig:psconfined} shows the dependence of the particle normal stresses on volume fraction for periodic suspensions and confined suspensions ($\chi = 0.1$, where $L_z$ = 20). The confinement level $\chi = 0.1$ can be regarded as {a} moderately confined regime. The general shapes of the particle normal stress for a confined suspension are similar to ones for a periodic suspension. Notably, the confinement appears to cause the opposite effects on the transverse and field directions. Compared to the periodic suspension, the transverse normal stresses ($\Sigma_{xx}^p, \Sigma_{yy}^p$) are lower, while the field normal stress ($\Sigma_{zz}^p$) is higher at all volume fractions considered. A difference of the particle stress between the periodic and confined suspensions gets larger as a volume fraction is increased and becomes constant in concentrated regimes.

Fig.~\ref{fig:ppconfined} compares the particle pressure between periodic and confined suspensions ($\chi = 0.1$). A difference of the particle pressure between the periodic and confined suspensions keeps larger as a volume fraction is increased. The noticeable change is that the confinement makes the particle pressure decrease and become negative earlier starting at $\phi \approx 20\%$. It implies that the confinement essentially results in augmenting the volumetric particle contraction. The short-range repulsive interactions between particles and boundaries could be responsible for this change, for which detailed investigations will be included in future work.

\begin{figure}
\centerline{
\includegraphics[width=3.4in]{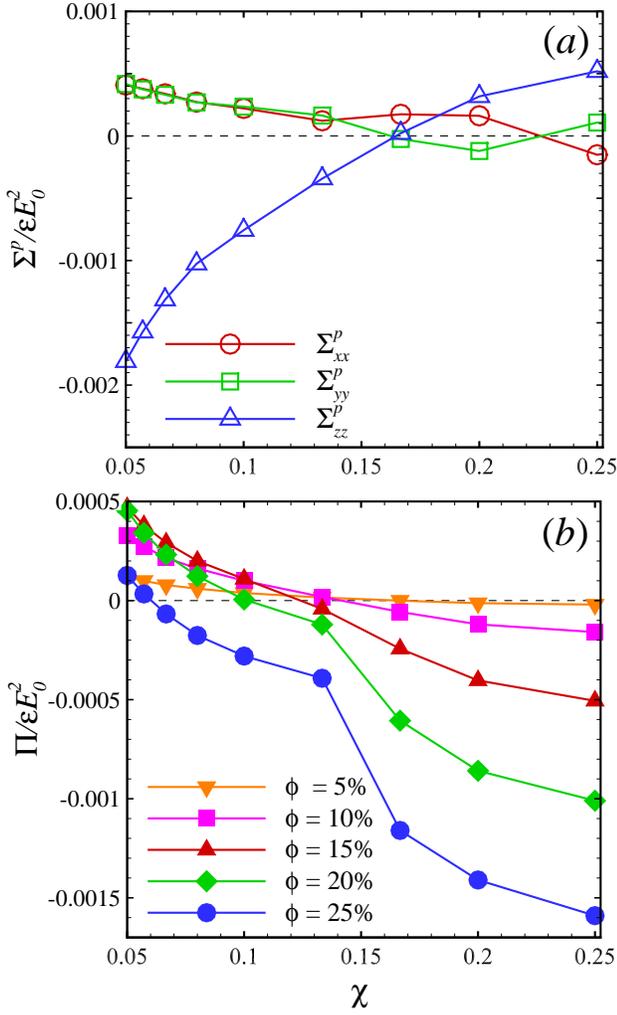}}% Here is how to import EPS art
\caption{\label{fig:cf} The effect of the level of confinement on ($a$) the particle normal stresses at $\phi = 10\%$ and ($b$) the particle pressure at different volume fractions. The level of confinement is captured by the confinement factor $\chi = 2a/L_z$, where $L_z$ is the electrode spacing.}
\end{figure} 

Lastly, the effects of the confinement factor $\chi$ on the particle stress and pressure are considered. Fig.~\ref{fig:cf}(a) shows the particle normal stresses of suspensions at $\phi = 10\%$ from weakly confined regime $\chi = 0.05$ to strongly confined regime $\chi = 0.25$ (note that $\chi = 0.05$ and $0.25$ correspond to 20 and 4 particle diameters, respectively). As the confinement level gets stronger (i.e., increasing the confinement factor), the noticeable rise of the normal stress in the field direction is observed, while a relatively small decrease is observed for the transverse directions. Thus, it appears that the change in the particle normal stress in the $z$ direction due to the increase in the level of confinement essentially governs the particle pressure reduction as no significant change is observed for the normal stresses in the $x$ and $y$ directions in the range of $\chi$. More interestingly, $\Sigma_{xx}^p$ and $\Sigma_{yy}^p$ are almost the same up to $\chi = 0.13$ but become separated from each other and almost opposite in sign beyond $\chi = 0.13$. This separation might result from the instability induced by the strong confinement. %(WHAT DO YOU MEAN HERE?).
% indicating a relative symmetry break in the particle normal stress with respect to $x$ and $y$-directions. 
Finally, the dependence of the particle pressure on the level of confinement is presented in Fig.~\ref{fig:cf}(b). Five different volume fractions up to $\phi = 25\%$ are considered because the corresponding particle pressures are all positive for a periodic domain, as seen in Fig.~\ref{fig:ppressure}. As the level of confinement increases, the particle pressure decreases with {the} confinement factor and then eventually becomes negative for all volume fractions. Moreover, the changeover at which the particle pressure turns into negative from positive arises earlier as a volume fraction increases. It is suggested that the strong confinement increases the volumetric contraction of the particle phase in a suspension. %The increase in particle concentration also appears to strengthen the negative pressure and thus decrease the critical confinement factor where the change to negative particle pressure takes place (see Fig.~\ref{fig:cf}(b)).
% Similar to the particle normal-stresses, the particle pressure seems to be saturated at very high confinement as the curve in Fig.~\ref{fig:cf}(b) reaches a plateau at strong $\chi$.

\section{Concluding remarks}

We have performed large-scale Stokesian dynamics simulations to study suspensions of ideally conductive spheres undergoing dipolophoresis, the combination of dielectrophoresis (DEP) and induced-charge electrophoresis (ICEP). {In the current system, it is found that ICEP dominates DEP as a result of the slower decay of hydrodynamic interactions -- a Stokes dipole for ICEP as $O(R^{-2})$ and a potential quadrupole for DEP as $O(R^{-4})$, where $R$ is the separation distance between two particles.} %It is found and ICEP dominates DEP in the current system as a result of the slower decay of interactions. 
%We previously studied the dynamics of such suspensions, where a non-trivial behavior was observed in concentrated regimes \cite{mirpark2019}. This behavior was reported to be correlated to change in the predominance of particle paring mechanism and transition in local microstructure.
The suspension dynamics was investigated by relaxation time, where it appears to relax very quickly with increasing volume fraction up to $\phi = 35\%$, but
% which is resulted from the 
%  which is explained by an increase in the magnitude of particle-particle interactions. % due to the reduction in characteristic separation distance. 
  %For the dilute suspension, the relaxation time inversely proportional to the square root of $\phi$, then it turns to linearly proportional to the inverse of volume fraction ($1/\phi$) at semi-dilute regime.
the relaxation time increases up to $\phi \approx 47.5\%$ and decreases again as approaching random {close} packing. %This non-monotonic behavior is also seen in a non-trivial behavior in suspension dynamics for the same system \cite{mirpark2019}. 
   %Therefore, the change in relaxation time at the concentrated suspension is suggested to be correlated with the change in the hydrodynamic diffusivity.
%   Additionally, the time-averaged interparticle gap $\bar{h}$ versus volume fraction exhibits four different regimes, including dilute, semi-dilute, concentrated, and very concentrated distinguished by different slopes in the log-log plot.
%    with a certain rate of change in $\bar{h}$ with $\phi$. 
The particle stress tensor is computed to characterize the suspension rheology and shows that the normal stresses are predominant over the shear stresses. The particle normal stress is {primarily} a sum of two contributions, the ICEP effect, which only contributes to the normal stress in the field direction, and hydrodynamic interactions, which contribute to both transverse and field directions.
 % based on Batchelor's calculation \cite{batchelor1970} of the average stress tensor,%

The non-monotonic behaviors of both particle normal stress ($\Sigma_{xx}^p, \Sigma_{yy}^p, \Sigma_{zz}^p$) and particle pressure ($\Pi =  -1/3\Sigma_{ii}^p$) with volume fraction were observed and shown to tie into the suspension dynamics and microstructure. At $\phi < 30\%$, the particle pressure $\Pi > 0$, which is similar to a hard-sphere suspension in a simple shear flow, implying that the particle pressure places the suspending fluid in tension relatively. Interestingly, the nature of particle pressure is changed at $\phi \approx 35\%$ at which it turns negative, indicating that the suspending fluid is now placed in compression. The negative particle pressure eventually becomes maximum at $\phi = 45\%$, where the hydrodynamic diffusivity and the velocity fluctuation reach a local maximum, as observed in our previous study \cite{mirpark2019}. It suggests that the negative particle pressure is connected to the observed non-trivial behaviors in the system.
  
Lastly, the wall confinement in the field direction is shown to strongly affect both particle normal stress and particle pressure. % in the field and transverse directions, respectively, although with no significant change in the general trend. 
It enhances the negative particle pressure drastically as a result of the augmentation of the volumetric particle contraction as a result of the short-range repulsive force between walls and particles. The confinement effect is further investigated by the confinement factor ($\chi = 2a/L_z$, where $L_z$ is the electrode spacing). It is observed that the field normal stress is more affected by the confinement than the transverse ones, and the particle pressure turns negative much earlier with increasing confinement factor. %The critical confinement factor, where the sign of particle pressure changes to negative, decreases with particle concentration seen to strengthen the negative particle pressure resulted from the confinement.
{For a confined suspension in microfluidic geometries, where the confinement is likely to introduce environmental heterogeneity, the conductive particles can display even more complex interactions with the walls, depending on the direction and frequency of the applied electric field, wall conductivity, and particle asymmetry \cite{kilic2011, gangwal2008, wu2009}. Such interactions have yet to be fully explained and will be a subject of interesting future work.}

Moving forward, the investigation of suspensions of ideally conductive particles in shear flow is necessary for a complete study of {the} rheology of such suspension, which will be included in future work. Another interesting future work is to control the rheology of such suspension via surface treatments such as surface coating or ion surface absorption. It was shown that a thin dielectric coating over ideally conductive particles leads to {the} change in the suspension dynamics from ICEP-dominated to DEP-dominated, where the local aggregation of particles is observed \cite{park2011d}. Thus, the surface treatments can be considered as a tuning factor to control dynamics and rheology of such suspension, which may provide a useful avenue toward new engineering applications. Finally, these distinct changing nature of rheological properties could suggest that different flow control techniques could be used to delay or promote the changing nature in such suspension system.

\begin{acknowledgments}
The authors gratefully acknowledge the financial support from the Collaboration Initiative and Interdisciplinary Research Grants at the University of Nebraska and in part from the National Science Foundation through a grant CBET-1936065 (Particulate and Multiphase Processes program). This work was completed utilizing the Holland Computing Center of the University of Nebraska, which receives support from the Nebraska Research Initiative.
\end{acknowledgments}

\section*{APPENDIX A: SPME ALGORITHM}
For large-scale dynamics driven by DEP and ICEP, we wish to calculate sums of form 
 \begin{align}\label{eq:us_p}
   &{\textbf{\textit{u}}}_s({\textbf{\textit{x}}}_{\alpha})=\sum_{\beta = 1}^{N} \mathbf{S}_P({\textbf{\textit{x}}}_{\beta}-{\textbf{\textit{x}}}_{\alpha}):\hat{\mathbf{z}}\hat{\mathbf{z}}, \\\label{eq:ut_p}
  & {\textbf{\textit{u}}}_t({\textbf{\textit{x}}}_{\alpha})=\sum_{\beta = 1}^{N} \mathbf{T}_P({\textbf{\textit{x}}}_{\beta}-{\textbf{\textit{x}}}_{\alpha}):\hat{\mathbf{z}}\hat{\mathbf{z}},
 \end{align}
where $\alpha=1,...,N$, and $\mathbf{S}_p$ and $\mathbf{T}_p$ denote the periodic versions of Green's function of Stoke dipole and a potential quadrupole. By making use of the known Ewald summation formula \citep{hasimoto1959,saintillan2005}, Eqs.~(\ref{eq:us_p}) and (\ref{eq:ut_p}) can be recast into the following Ewald summations:
\begin{equation}\label{eq:us_ewald}
\begin{aligned}
      {\textbf{\textit{u}}}_s({\textbf{\textit{x}}}_{\alpha}) = \sum_{\textbf{\textit{p}}}\sum_{\beta = 1}^{N}\mathbf{A}_s(\xi,{\textbf{\textit{x}}}_{\beta}-{\textbf{\textit{x}}}_{\alpha}+{\textbf{\textit{p}}}):\hat{\mathbf{z}}\hat{\mathbf{z}} \ + \\ \sum_{k\neq 0}e^{-2\pi i {\textbf{\textit{k}}} \cdot {\textbf{\textit{x}}}_{\alpha}}S({\textbf{\textit{k}}})\mathbf{B}_s(\xi,{\textbf{\textit{k}}}):\hat{\mathbf{z}}\hat{\mathbf{z}},
\end{aligned}
\end{equation}
\begin{equation}\label{eq:ut_ewald}
\begin{aligned}
      {\textbf{\textit{u}}}_t({\textbf{\textit{x}}}_{\alpha}) = \sum_{\textbf{\textit{p}}}\sum_{\beta = 1}^{N}\mathbf{A}_t(\xi,{\textbf{\textit{x}}}_{\beta}-{\textbf{\textit{x}}}_{\alpha}+{\textbf{\textit{p}}}):\hat{\mathbf{z}}\hat{\mathbf{z}} \ + \\ \sum_{k\neq 0}e^{-2\pi i {\textbf{\textit{k}}} \cdot {\textbf{\textit{x}}}_{\alpha}}S({\textbf{\textit{k}}})\mathbf{B}_t(\xi,{\textbf{\textit{k}}}):\hat{\mathbf{z}}\hat{\mathbf{z}},
\end{aligned}
\end{equation}
where $\xi$, called the Ewald coefficient, which determines the relative importance of the real and Fourier sums. This coefficient is user-defined and is chosen to minimize the overall cost of the algorithm. The first sums (real sums) in Eqs.~(\ref{eq:us_ewald}) and (\ref{eq:ut_ewald}) are over all particle positions ${\textbf{\textit{x}}}_{\beta}$ and their periodic images (which are denoted by the lattice vectors ${\textbf{\textit{p}}}$), and the second sums (Fourier sums) are over wave vector ${\textbf{\textit{k}}}$. The structure factor $S({\textbf{\textit{k}}})$ in Eqs.~(\ref{eq:us_ewald}) and (\ref{eq:ut_ewald}) is obtained for the suspension as follows
\begin{equation}\label{eq:s_k}
  S({\textbf{\textit{k}}}) =  \sum_{\beta = 1}^{N}e^{2\pi i {\textbf{\textit{k}}} \cdot {\textbf{\textit{x}}}_{\beta}}.
\end{equation}
The convolution kernels $\mathbf{A}_s$, $\mathbf{A}_t$, $\mathbf{B}_s$ and $\mathbf{B}_t$ are third-order tensors and can be calculated analytically (see Park \& Saintillan \cite{park2010}). These tenors decay exponentially, which consequently results in exponential convergence of the sums in Eqs.~(\ref{eq:us_ewald}) and (\ref{eq:ut_ewald}). The details of evaluating these tenors in SPME algorithm are provided in Saintillan \textit{et al.} \cite{saintillan2005}. 
An $O(N)$ cost for the evaluation of the real sums at all particle positions is obtainable by choosing Ewald coefficient $\xi$ so as to exclude all the terms beyond a fixed cutoff distance $r_c$ from the reals sums, which allows truncation of these sums after a finite number of terms independent of the system size. For the Fourier sums (second sums), the particle distribution is transformed to Fourier space using the fast Fourier transform algorithm, after assigning to Cartesian gird by B-spline interpolation \cite{essmann1995}. It yields structure factor $S({\textbf{\textit{k}}})$, which is then multiplied by the convolution kernels $\mathbf{B}_s$ and $\mathbf{B}_t$. The inverse Fourier transform is applied, and the value of Fourier sums is determined at the particle locations by interpolation.
 The computation cost of Fourier sums is limited by the fast Fourier transform algorithm, scaling as $O(K \textrm{log}K)$ with the number $K$ of grid points (or Fourier modes). This number is typically chosen to be proportional to the number of particles $N$; therefore, the overall cost for the velocities evaluation is $O(N\textrm{log}N)$.

%\appendix

%\section{Appendixes}

%To start the appendixes, use the \verb+\appendix+ command.
%This signals that all following section commands refer to %appendixes
%instead of regular sections. Therefore, the \verb+\appendix+ %command
%should be used only once---to set up the section commands to act as
%appendixes. Thereafter normal section commands are used. The %heading
%for a section can be left empty. For example,
%\begin{verbatim}
%\appendix
%\section{}
%\end{verbatim}
%will produce an appendix heading that says ``APPENDIX A'' and
%\begin{verbatim}
%\appendix
%\section{Background}
%\end{verbatim}
%will produce an appendix heading that says ``APPENDIX A: %BACKGROUND''
%(note that the colon is set automatically).

%If there is only one appendix, then the letter ``A'' should not
%appear. This is suppressed by using the star version of the %appendix
%command (\verb+\appendix*+ in the place of \verb+\appendix+).

%\section{A little more on appendixes}

%Observe that this appendix was started by using
%\begin{verbatim}
%\section{A little more on appendixes}
%\end{verbatim}

%Note the equation number in an appendix:
%\begin{equation}
%E=mc^2.
%\end{equation}

%\subsection{\label{app:subsec}A subsection in an appendix}

%You can use a subsection or subsubsection in an appendix. Note the
%numbering: we are now in Appendix~\ref{app:subsec}.

%\subsubsection{\label{app:subsubsec}A subsubsection in an appendix}

\nocite{*}
\setstretch{1.5}
\section*{REFERENCES}
\makeatletter
\renewcommand{\@biblabel}[1]{[#1]\ \ }
\makeatother
\bibliography{aipsamp}% Produces the bibliography via BibTeX.
%\bibpunct{(}{)}{;}{n}{}{,}

\end{document}